\magnification=1200 
\input epsf 

\def\th{\theta}
 
\def\Dl{\Delta} 

\def\Om{\Omega}
\def\omt{\Omega_T}
\def\omo{\Omega_o}
\def\omc{\Omega_{-1}}
\def\oms{\Omega_{-1/3}}
\def\omr{\Omega_{1/3}}
\def\zd{z_{dec}}
\def\sg{\sigma} 
\def\dl{\delta}
\def\gm{\gamma}
\def\Gm{\Gamma}

\rightline{$\,$}
\medskip 
\centerline{\bf Acoustic Peak Spacing, Cosmological Density, and Equation 
of State} 
\bigskip 
\centerline{Eric V.~Linder} 
\centerline{University of Massachusetts, Department of Physics and Astronomy, 
Amherst, MA 01003}
\bigskip 
\centerline{ABSTRACT}
\medskip 
The spacing of the acoustic peaks in the cosmic microwave background 
radiation anisotropy multipole spectrum has been claimed to provide 
the value of the total cosmological density overtly, ``written on the 
sky.''  Through a semianalytic analysis of the cosmological evolution 
of the sound horizon and the physics of decoupling we address the 
robustness of the relation between the peak spacing and the cosmological 
density.  In fact, the asymptotic distance and horizon scalings often 
used are not good approximations, and the individual densities and 
equations of state of different components do enter the problem.   
An observed spacing could be fit by models with 
different total densities.  We investigate the different regions of 
density-equation of state parameter space and also provide accurate fitting 
formulas for the peak spacing as a function of matter density, total density, 
and additional component equation of state (e.g.~cosmological constant or 
cosmic strings).  Limits provided by peak spacing measurements on the 
number of neutrino species and the baryon-photon ratio are also addressed. 

\bigskip 
\centerline{\bf 1. Introduction}
\medskip 
The cosmic microwave background (CMB) radiation provides a cornucopia of 
cosmological information about the early universe and the formation 
of large scale structure.  With the increasing sensitivity of medium scale 
angular anisotropy experiments, both ground based and the near future 
space missions MAP and Planck, further information on the properties of 
our universe is imminent.  Measurements on scales of 10-60 arcminutes  
cover the horizon scales around the epochs of recombination and last 
scattering and promise a wealth of data on the large scale properties 
of our universe as well as indications of the early universe origin of 
fluctuations that become cosmological structure. 

In particular, they hold the hope of measuring the total cosmological 
energy density and hence elucidating the ultimate fate of our universe, 
as well as providing evidence on the validity of the inflationary picture 
of perturbation generation.  Such a lofty goal would be reached through 
a series of measurements using straightforward, well understood, physical 
theory, and would avoid more tortuous and uncertain paths such as the 
distance ladder and the nonlinear evolution of complex galaxies and stars.  

With the prospect of the revelatory data looming, it behooves us to 
analyze carefully the extent to which the translation from peak multipole 
spacing measurements to the single number of the total energy density of 
the universe is indeed obvious and clean.  At first sight it seems 
astonishing that despite having different species of cosmological 
inhabitants -- photons, baryons, neutrinos, cold dark matter, cosmological 
constant, etc.~-- and an uncertain expansion rate (Hubble constant) that 
the spacing should depend on this one single parameter.  Fortunately, the 
physics of baryon-photon decoupling and cosmological distances is 
sufficiently simple that we can go a long way through analytic 
investigation.  Of course, numerical calculations of not only the peak spacing 
but their amplitudes and the full multipole spectrum of microwave background 
anisotropies have been carried out (e.g.~Zaldarriaga, Seljak, \& 
Bertschinger 1997; Kamionkowski, Spergel, \& Sugiyama 1994), and a  
rigorous derivation of cosmological parameter values should rely on the full 
information and accuracy they possess (e.g.~Zaldarriaga, Spergel, \& Seljak 
1997; Bond, Efstathiou, \& Tegmark 1997). 
But for understanding the physical origin of the results in different regions 
of parameter space, this analytic exploration provides a heuristic and 
surprisingly accurate portrait. 

\bigskip
\centerline{\bf 2. Acoustic Oscillation Anisotropies}
\medskip 
On medium angular scales, $\th\approx 10'-2^\circ$ or multipoles 
$l\approx 100-1000$, the dominant anisotropy in the CMB is expected 
to be acoustic, or Doppler, oscillations (Hu, Sugiyama, \& Silk 1997; 
Bond et al.~1994; occasionally these are called Sakharov (1966) 
oscillations).  
These are due to harmonic density perturbations in the coupled 
baryon-photon fluid in the prerecombination epoch.  While after recombination 
these will grow to form the large scale structure of the universe, the 
density perturbations 
cannot grow as long as the matter is tightly coupled to the radiation, 
e.g.~through Thomson scattering.  This 
situation persists even after the end of the radiation dominated era 
due to the high entropy, or low baryon to photon ratio.  Thus even 
when the matter energy density governs the expansion rate of the universe,  
the coupling forces the behavior to be oscillational -- a sound or acoustic 
wave. 

As the universe expands, the Thomson scattering rate drops below the 
cosmological expansion rate and the coupling reactions become inefficient. 
Matter recombines to neutral atoms and the ionization fraction freezes 
out.  The photons become free streaming and the lack of further interaction 
preserves the density irregularities imprinted on the photon field by the 
matter oscillations.  These appear as angular anisotropies, with the 
largest angular scale arising from the largest matter wavelength $\lambda$ 
-- that of the sound horizon at decoupling. 

In CMB studies one often works in multipole space $l\sim \th^{-1}\sim 
\lambda^{-1}$. A length scale translates to 
an angle $\th$, or multipole $l$, through the angular diameter distance 
$r_a$ by $l=kr_a=2\pi r_a/\lambda=\pi/\th$, where $k$ is the wavenumber.  
Thus one expects a harmonic 
series of anisotropies corresponding to acoustic normal modes, with peaks 
at $l_1:l_2:l_3\dots\approx 1:2:3\dots$.  The peak spacing $\Dl l$ is 
a direct measure of the maximum wavelength, the horizon diameter 
$\lambda=2r_h$.  Thus, 
$$
\Dl l=\pi r_a/r_h.\eqno(1) 
$$ 

In attempting to learn about cosmological parameters by observing the 
anisotropy acoustic peak spacing or locations, we are simply applying 
the classical angular diameter distance test to the CMB.  This has many 
virtues over its use with galaxies and other conventional objects, 
including astrophysical simplicity and physical understanding of the 
source, lack of evolutionary or selection effects, and the high redshift 
improving the discrimination between cosmological models, as advocated 
by Linder (1988a). 

In order to compare observations of the multipole spacing to cosmological 
models we need three quantities: the angular diameter distance 
$r_a$ out to redshift $z$, the sound horizon scale $r_h$ at redshift $z$, 
and the redshift of decoupling $\zd$, at which we evaluate these functions.  
Each depends on numerous subsidiary cosmological variables.  Before 
looking at these dependences let us note a few issues regarding the 
applicability of this acoustic oscillation picture. 

The ansatz relies on the presence of harmonic perturbations in the coupled 
baryon-photon fluid prior to decoupling.  While this is a generic feature of 
the adiabatic fluctuations generated in inflationary scenarios, it does not 
have to exist generally.  The presence of isocurvature perturbations, such 
as those arising from topological defect models, suppresses some of the peaks 
and changes the peak spacing (see Hu \& White 1996).  It is also possible 
to arrange active, causal processes in such a way as to create peaks that 
mimic the results of inflation (Turok 1996).  

Moreover, although in a 
purely harmonic mode the first peak occurs at multipole $l_1=\Dl l$ and 
the spacing $l_{m+1}-l_m=\Dl l$ for all $m$, not just $m\gg1$, 
gravitational forcing terms in the oscillations slightly change these 
predictions (Hu \& White 1996).  Given a theory for the origin of the 
perturbations, however, the locations and small $m$ spacings (they reach 
the asymptotic value by $m\approx3$) can be related to the pure harmonic 
value $\Dl l$.  Thus, though we phrase all our results in terms of the 
spacing $\Dl l$, they can be applied as well to the peak locations. 

Although adiabatic perturbations arise naturally within inflationary 
theory, we will not restrict ourselves to flat cosmological models, 
both because of the possibility of open inflationary universes, 
and because adiabatic perturbations can be generated via other, causal 
processes.  With these points in mind, this paper operates within the 
picture of the acoustic peaks directly tracing the sound horizon scale, 
for however far that is valid. 

\bigskip
\centerline{\bf 3. Physical Variables} 
\medskip 
As illustrated in (1), the peak multipole spacing depends on two 
distance scales and the redshift at which they are evaluated.  The 
distance-redshift relations are straightforward cosmological 
expressions.  The redshift is that of decoupling, for this is when 
the physical imprinting of the acoustic anisotropies in the CMB 
temperature pattern occurs, when the photons become unaffected by 
further interactions with the matter. 

\bigskip 
\leftline{\bf 3a. Angular Diameter Distance}
\medskip 
In a homogeneous and isotropic universe, i.e.~a Friedmann-Robertson-Walker 
(FRW) model, the angular diameter distance is given by 
$$
r_a(z)=(1+z)^{-1}(1-\omt)^{-1/2}\sinh \left[(1-\omt)^{1/2}\int_1^{1+z} 
dy/H(y)\right],\eqno(2)
$$
where $\omt$ is the total cosmological energy density in units of the 
critical density $\rho_c=3H^2(0)/8\pi$, $y=1+z$, and $H(y)$ is the 
Hubble parameter, 
$$
H(y)=H(0)\left[\sum_\sg \Om_\sg y^{3(1+\sg)}+(1-\omt)y^2\right]^{1/2}. 
\eqno(3)
$$
(See Linder 1988a,b, 1997 for general distance-redshift and cosmological 
parameter expressions for multiple components and equations of state.) 
The equation of state of a cosmological component is given by 
$\sg=p/\rho$, where $p$ is its pressure, $\rho$ its energy density, and 
$\Om_\sg$ its dimensionless density $\rho/\rho_c$. 

Note that $r_a$ depends not only on $\omt=\sum \Om_\sg$ but also on each 
component individually, though generally one particular equation of state 
will dominate at a given redshift.  One subtlety involves the use of the 
FRW model despite the clumpiness of matter -- i.e.~structure -- along the 
line of sight, at least at low redshift.  The question of when to use clumpy 
model distances instead of FRW distances is not wholly settled, but 
according to recent criteria of Linder (1998), FRW relations can be applied 
with confidence when considering angular scales $\th\gg 5''$ ($l\ll 10^5$), 
as we do. 

\bigskip
\leftline{\bf 3b. Sound Horizon Scale}
\medskip 
The particle horizon distance is $r_p=a\int dt/a$, where $a(t)$ is the 
expansion parameter, or scale factor, and $t$ is the time.  To obtain the 
sound horizon we simply correct this for the speed of propagation, or 
sound speed $c_s$, of the acoustic waves: 
$$
r_h(z)=(1+z)^{-1}\int_{1+z}^\infty dy\,c_s(y)/H(y),\eqno(4)
$$
since $H(y)=a^{-1}da/dt$ and $y=a^{-1}$. 

The sound speed is closely related to the equation of state parameter 
$\sg$: $c_s=\sqrt{dp/d\rho}$ while $\sg=p/\rho$.  At the epoch of decoupling 
and 
earlier we expect the only two significant components to be nonrelativistic 
(pressureless) matter, $\sg\approx (1/3)v^2/c^2\approx 0$, and relativistic 
particles -- photons and neutrinos -- with $\sg=1/3$.  The only acoustic 
coupling that exists, however, is between the baryons and photons.  Thus 
the total pressure fluctuation in the acoustic medium is 
$dp=dp_\gm=(1/3)d\rho_\gm$ and 
the total energy density fluctuation is $d\rho=d\rho_\gamma+d\rho_b$. 

For adiabatic perturbations the total entropy fluctuation must vanish; 
since the entropy is directly proportional to the baryon-photon number 
ratio $\eta=n_b/n_\gm$, this implies $(\dl n/n)_b=(\dl n/n)_\gm$. 
For matter, $(\dl n/n)_b=(\dl\rho/\rho)_b$ while for blackbody radiation 
the number density goes as the temperature $T^3$ and the energy density 
goes as $T^4$, so $(\dl n/n)_\gm=(3/4)(\dl\rho/\rho)_\gm$.  Thus, 
$\dl\rho_b=(3/4)(\rho_b/\rho_\gm)\,\dl\rho_\gm$.  Finally, the evolution 
with expansion of the two components is $\rho_b\sim y^3$ and $\rho_\gm\sim 
y^4$, so we find 
$$
c_s(y)=\sqrt{dp/d\rho}=\sqrt{{1\over 3}\,{1\over 1+(3/4)(\Om_b/ 
\Om_\gm)\,y^{-1}}}\,.\eqno(5)
$$

Defining a redshift of baryon-photon equality, $y_{eq(b,\gm)}=\Om_b/ 
\Om_\gm$, we see that for $y\gg y_{eq(b,\gm)}$, when photons dominate, 
the effective equation of state is $\sg_{eff}\equiv c_s^2=1/3$, while as 
the universe expands and 
baryons come to dominate for $y\ll y_{eq(b,\gm)}$, $\sg_{eff}\to 0$.  Thus 
the sound speed varies betweens $1/\sqrt{3}$ and 0, decreasing as 
baryons become more important (this can be viewed as increasing the 
effective mass of the acoustic phonon -- see Hu, Sugiyama, \& Silk 1997).  
Note that the redshift of baryon-photon equality is not the same as the 
redshift of matter-radiation equality, unless all matter is baryonic 
(no cold dark matter, for example), and all radiation is photons (e.g.~no 
light or massless neutrinos). 

\bigskip 
\leftline{\bf 3c. Decoupling Redshift} 
\medskip 
The last ingredient necessary is the redshift at which to evaluate the 
two distance scales, that of decoupling when the photons receive their 
final anisotropy imprint.  The coupling process is dominated by Thomson 
scattering between the photons and the free electrons of the perennially 
ionized hydrogen and proceeds at a time dependent reaction rate $\Gamma(z)
=\sg_Tn_e(z)$, where $\sg_T$ is the Thomson cross section and $n_e$ the 
electron number density.  

As the universe expands, the photon gas cools, 
eventually dropping below the ionization energy of hydrogen, leading to 
recombination of the electrons into atoms.  Due to the high entropy, or 
low baryon-photon number $\eta$, this actually occurs when the photon 
temperature is $2\ln \eta\approx 1/40$ of the hydrogen ground state 
ionization energy.  However, although related, the decoupling of the photons 
and baryons is a slightly different situation: what is important here is 
the number of interactions ongoing between the two components.  When the 
Thomson reaction rate drops below the expansion rate $H\sim t^{-1}$ of 
the universe, or equivalently the photon mean free path exceeds the light 
horizon, then interactions effectively cease: $N_{int}=\int_{t_{dec}}^\infty 
\Gamma\,dt <1$. 

The criteria for decoupling, therefore, is $\Gm\le H$ (cf.~Kolb \& Turner 
1990).  Both rates are redshift dependent: as the universe expands (redshift 
decreases) the electron density is diluted and also the Hubble parameter 
decreases.  Setting $\Gm=H$ defines the desired redshift of decoupling, 
$z_{dec}$.  To find the electron density $n_e=X_e\eta n_\gm(T_\gm)$ in 
$\Gm$ one cannot simply use the Saha expression for the ionization fraction 
$X_e=n_e/n_b$ because the atomic levels are not in statistical equilibrium. 
Jones \& Wyse (1985) provide an excellent analysis of the relevant physics 
and applicable approximations.  We follow their work, generalizing it 
slightly.  By comparing various atomic excitation rates, they reduce the 
expression for the ionization fraction as a function of redshift to a 
Riccati equation and find that near decoupling a quasiequilibrium WKBJ 
solution obtains, similar to the analysis for the nucleosynthesis rate 
equations done by Esmailzadeh, Starkman, \& Dimopoulos (1991). 

This fixed point solution gives $X_e\sim(\omt h^2)^{1/2}f^{-1}(z,\omt)\,
(\Om_b h^2)^{-1}$, where $h=H(0)/$(100 km s$^{-1}$ Mpc$^{-1}$) and 
$f(z,\omt)$ involves the ratio of the decoupling redshift to the redshift 
of equality of energy density in the matter and radiation, $z_{eq}$.  
Retracing their analysis for a general, not just dust (zero pressure), 
universe, i.e.~allowing arbitrary equation of state components with density 
$\Om_\sg$, reveals that the factor $(\omt h^2)^{1/2}f^{-1}(z,\omt)$ arises 
from $(dt/dz)^{-1}\sim H(z)$ and the $\Om_b h^2$ factor comes from $n_b$, 
since the total recombination rate is proportional to $n_en_p\sim n_b^2$ 
while the competing total ionization rate is proportional to $n_b$ ($n_e$, 
$n_p$ are 
the number densities of electrons and protons, respectively).  Therefore 
the generalized decoupling condition $\Gm=H$ is 
$$
\eqalign{\Gm(z_{dec})&=\sg_T n_e\sim X_en_b\sim H(z)n_b^{-1}n_b=H(z_{dec})\cr 
&\Rightarrow\quad z_{dec}=z_{dec}(T_\gm),\cr}\eqno(6)
$$
i.e.~$z_{dec}$ is practically a constant (except that the atomic transitions 
also depend on the background temperature $T_\gamma$). 

One has the remarkable conclusion that cancellations in $H(z)$ and $n_b$ 
lead to the redshift of decoupling being extraordinarily insensitive to 
the baryon content of the universe, the total density and density of 
individual components, and the overall expansion rate.  The redshift is 
almost completely determined by the photon temperature, which is of course 
fixed by CMB measurements, and implies $z_{dec}=1100$.  This is the value at 
which we evaluate the distances $r_a$ and $r_h$ for all models. 


Jones \& Wyse (1985) carefully examine the approximations they use and 
these appear robust.  We will note that the leading correction term (the 
last term in their equation A13) is proportional to $\Om_b/\omt$ and so 
the approximation should be especially good for universes with $\Om_b\ll 
\omt$ (as they state; also see underived equation 3.101 of Kolb \& Turner 
1990) -- this holds for all models we consider. 

\medskip
\leftline{\bf 3d. Summary of Cosmological Dependences}
\medskip 
Before calculating $r_a(z_{dec})$ and $r_h(\zd)$ to obtain the multipole 
peak spacing we pause to review the cosmological ingredients that enter 
into the problem.  In broad terms: $r_a$ involves $\omt$ and $H(z)$; 
$r_h$ involves $c_s(z)$ and $H(z)$; $\zd$ involves $\Gm(z)$ and $H(z)$. 
Each of these major variables, however, has subsidiary elements entering: 

$\bullet$\quad $H(z)$ -- component densities $\Om_\sg$ and equations of 
state $\sg$; current value $h=H(0)/$(100 km s$^{-1}$ Mpc$^{-1}$) 

$\bullet$\quad $c_s(z)$ -- baryon-photon ratio $\eta$; helium abundance 
$Y$ through translation of $n_b$ to $\Om_b$ (i.e.~average mass of a 
baryonic particle: see Section 5b) 

$\bullet$\quad $\Om_{1/3}$ -- number of neutrino species $N_\nu$; Hubble 
constant $h$ affects translation of $T_\gm$ to $\Om_{1/3}\sim h^{-2}T_
\gm^4$; $\Om_{1/3}$ enters into $H(z)$ and definition of $z_{eq}=\Om_
{1/3}/\Om_o$ 

$\bullet$\quad $r_a(z)$ -- clumpiness of density distribution, as discussed 
in Section 3a 
\smallskip 
We keep these in mind for two reasons: 1) to understand the original claims 
that the peak spacing $\Dl l$ was degenerate in these ingredients and 
depended only on $\omt$, 
and 2) to attempt to use the wealth and accuracy of CMB anisotropy data to 
probe these additional cosmological quantities when we see how the putative 
degeneracy breaks down.  After obtaining results for $\Dl l$ in the next 
section we return to this question in Section 5. 

\bigskip 
\centerline{\bf 4. Peak Spacing} 
\medskip 
Using equations (1)-(5) and $\zd=1100$ we obtain analytic or quadrature 
expressions for the anisotropy multipole peak spacing $\Dl l$.  We examine 
open and flat models, $\omt\le1$, with three cases of components: pure 
dust ($\sg=0$) with density $\Om_o$; dust plus a cosmological constant 
($\sg=-1$) 
with densities $\Om_o$, $\Om_{-1}$; dust plus a $\sg=-1/3$ component with 
densities $\omo$, $\oms$.  [$\oms$ does {\it not} include curvature energy 
$1-\omt$; 
it is a distinct component with $p=-(1/3)\rho$, corresponding perhaps to 
a cosmic string network.  Its function is simply to illustrate the effect 
of the equation of state on the anisotropy multipole pattern.  It is also 
interesting to consider because it doesn't affect the kinematics of the 
universe; that is, a universe with $\omt=\omo+\oms$ has the same $H(z)$ 
as one with just $\omt'=\omo$ (Linder 1988b).]   

We do not 
restrict to the flat inflationary case of $\omo+\omc=1$.  We take $T_\gm
=2.73\,K$, i.e.~$\Om_\gm=2.5\times 10^{-5}h^{-2}$, and three neutrino 
species so $\omr=4.2\times 10^{-5}h^{-2}$.  We ignore $Y$, take 
$\Om_bh^2=0.0125$, and consider $h=0.5$ and $h=1$ cases.  Effects of 
changing these parameters are discussed in Section 5b. 

First, let us explore the early argument that $\Dl l$ is degenerate with 
respect to all parameters but $\omt$.  Initially consider a dust 
universe.  At asymptotically high redshift, the angular diameter distance 
$r_a\approx 2H^{-1}(0)\omo^{-1}z^{-1}$, as can be seen by manipulation of 
(2) into the Mattig (1958; more accessible in Linder 1988b, eqs.~30, A4) 
analytic form.  Similarly, from (4) the sound 
horizon $r_h\to (2/\sqrt{3})H^{-1}(0)\omo^{-1/2}z^{-3/2}$ since the baryon 
loading is negligible at very high redshift.  Thus $\Dl l\to \pi\sqrt{3}\, 
z^{1/2}\omo^{-1/2}$.  If one assumes that $\zd=1100$ is in the asymptotic 
regime then one would obtain $\Dl l\approx 180\omo^{-1/2}$ 
(i.e.~$\theta\approx 1^\circ\times\omo^{1/2}$).  For a flat inflationary 
universe, 
$\omo+\omc=1$, $r_h$ is little affected and one asymptotically has 
$r_a\sim\omo^{-1/2}$ so $\Dl l\sim\omo^0$ -- independent of the individual 
components. 

This sort of estimation led to the early belief that the peak 
spacing depended only on $\omt$, not $\omo$ and $\omc$ individually (of 
course this overlooked the point that by assuming a flat universe one 
already knew $\omt$ -- what we supposedly were trying to find from the 
CMB acoustic anisotropy location). 

However, one is in fact not in the asymptotic regime.  In the dust case, 
the leading correction 
to $r_a$ is of order $(\omo z)^{-1/2}$, which can be as large as 10\%, the 
influence of the radiation component on $H(z)$ is not negligible for $r_h$ 
since $z_{eq}/\zd\ge(1/20)$ (it is a fairly small effect for $r_a$, $\le3\%$), 
and the baryon loading effect in $c_s$ is at maximum of order $(3/4)z_{eq
(b,\gm)}/\zd\approx 34\%$ (smaller when integrated over $z>\zd$).  One 
actually finds $r_a\sim\omo^{-0.93}$ and $r_h\sim\omo^{-0.32}$ so 
$\Dl l\sim \omo^{-0.6}$.  Remarkably, for a flat inflationary universe 
$\omo+\omc=1$, $r_a\sim\omo^{-0.40}$, $r_h\sim\omo^{-0.32}$ so $\Dl l\sim 
\omo^{-0.08}$ -- indeed rather insensitive to the individual component 
densities.  This breaks down, however, for nonflat universes. 

In the past few years numerous numerical computations (e.g.~Zaldarriaga, 
Seljak, \& Bertschinger 1997; Bond, Efstathiou, \& Tegmark; Kamionkowski, 
Spergel, \& Sugiyama 1994), properly including the effects discussed here, 
have calculated the multipole spectrum including 
the peak spacings.  However, no systematic analysis (beyond the useful 
animations of Hu -- see 
http://www.sns.ias.edu/\~{}whu/physics/physics.html) 
has been made of the physical scalings, especially 
for open models, providing simple analytical fits to the parameter 
dependences and demonstrating the role of individual component densities and 
equations of state.  Indeed, the myth persists within large sections of the 
astrophysical community that one or two imminent measurements of the first 
few Doppler (acoustic) peak locations will fix the total density and hence 
fate of the universe.  Here we attempt to portray the results almost as 
accurately as the matter Boltzmann--radiative transfer numerical simulations 
but with 
explicit fits and scaling of physical dependences, to illustrate the true 
blend of simplicity and complexity in the results. 

Figures 1 show the multipole peak spacing for the various models.  Note 
that the vertical spread of the curves demonstrates the lack of degeneracy 
with respect to the individual components -- i.e.~it is not just $\omt$ 
that defines $\Dl l$.  We do see that flat models have a small dispersion, 
as previously mentioned.  The bottom dashed curve gives the naive, often 
quoted $\omt^{-1/2}$ dependence; this is never an acceptable fit.  The 
bottom solid curve gives a derived $\omt^{-0.59}$ ($\omt^{-0.69}$ for 
$h=0.5$) fit for pure dust ($\omo=\omt$) universes.  Clearly the presence 
of individual components does significantly affect the peak spacing. 
Conversely, observations of acoustic peak locations does not uniquely 
determine the total density of the universe.  This is discussed further 
in Section 5a. 

\topinsert 
\vskip-1.0truein
\centerline{
\epsfxsize=7.4truein
\epsfbox{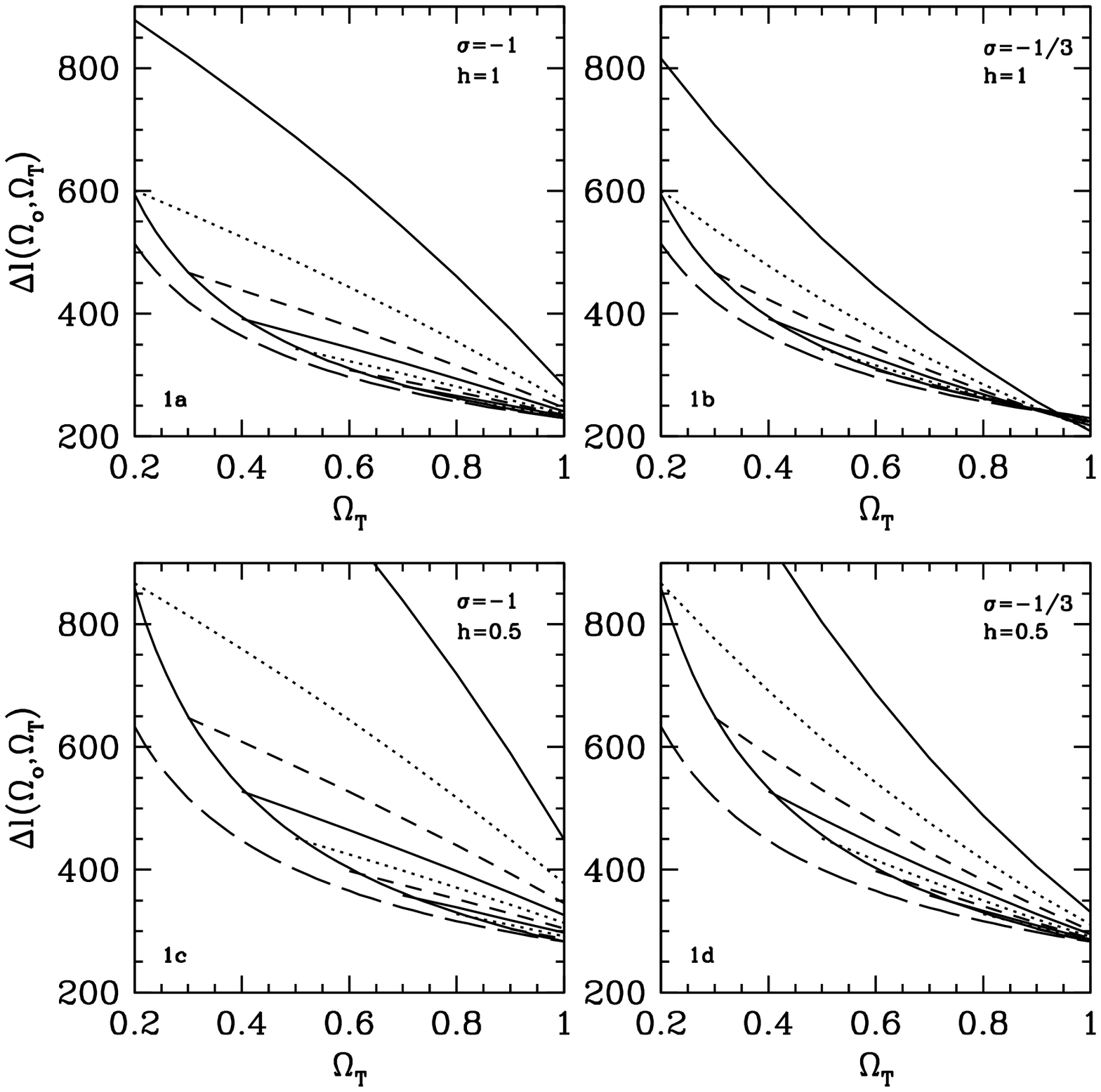}
}
\vskip-2.5truein\nobreak 
\noindent {\bf Figure 1:}\quad 
The peak multipole spacing $\Dl l$ is plotted 
against the total cosmological density $\omt$ for a variety of models. 
Each panel is labelled by the component equation of state $\sg$ (all 
include dust as well) and Hubble constant $h$.  The curves from the top 
down in each panel are for constant $\omo$, starting from $\omo=0.1$ 
(solid), $\omo=0.2$ (dotted), $\omo=0.3$ (dashed), etc.  They run for 
values of $\omt\ge\omo$.  The bottom two curves are fits for pure dust 
models (where $\omt=\omo$), serving as the lower envelope.  The solid 
bottom curve shows the fit of this paper (see equation 7 and Table 1) while 
the bottom dashed curve gives the naive asymptotic fit proportional to 
$\omt^{-1/2}$. 
\endinsert 

As the equation of state parameter $\sg$ of the second component approaches 
zero (that of the first component, dust), the dispersion at fixed $\omt$ 
lessens, as expected.  But there is no qualitative, smoking gun difference 
between adding different equation of state components, no obvious 
observational signature.  [The pure dust results ($\omo=\omt$) are of course 
the same in Figures 1a and 1b, and 1c and 1d.] 

Decreasing the Hubble constant has a large effect, increasing both the 
values and dispersion of the peak spacings.  Again this is in contradiction 
to the accepted asymptotic picture, that held that the Hubble constant only 
entered into the heights and not locations of the anisotropy peaks.  
(Rigorously, $\Dl l$ would be independent of $h$ only for flat 
universes where $z_{eq}\gg\zd$.)  Numerical 
simulations, however, show the same dependence exhibited here; for example 
Figure 2 of Hu \& White (1997; also see Figure 3) finds 
$\Dl l(\omo=1,\,\omt=1)_{h=0.5} 
\approx285$, in excellent agreement with our value of 283.  Note that the 
dashed asymptotic behavior $\omt^{-1/2}$ is here normalized to $\Dl l(1,1)$ 
for each $h$; if we used the true naive approach of $\Dl l(1,1)$ 
being independent of $h$, the fit of this curve to the spacings would be 
even worse. 

Analytic fits to $\Dl 
l$ incorporating scaling with the relevant physical variables: the matter 
density $\omo$, total density $\omt$, and Hubble constant $h$ are useful 
in presenting the essential physics.  Of course 
in the old $\omt^{-1/2}$ fit neither $\omo$ nor $h$ would appear.  The 
general form we derive is 
$$
\Dl l(\omo,\omt)/\Dl l(1,1)=\omo^{-x}+\left(\omo^{-y}-\omo^{-x}\right)\,
\left({{1-\omt}\over {1-\omo}}\right)^z.\eqno(7)
$$
Two particular cases of interest are the flat (possibly inflationary) case 
when $\omt=1$: here $\Dl l\sim\omo^{-x}$, and the pure dust case, 
$\omo=\omt$, with $\Dl l\sim \omo^{-y}$.  Table 1 gives the values $x,y,z$ 
for the equations of state and Hubble constants adopted.  Equation (7) is 
a central result of this paper and provides an excellent fit to the results 
for $\Dl l$ plotted in Figures 1, obtained from solving equations (1)-(5). 
The approximations are good to about 1\% rms with maximum deviation about 
2\%, and thus serve as useful and accurate guides to the physical 
dependences, short of carrying out the full numerical generation of the 
CMB multipole spectrum. 

\medskip 
\centerline{TABLE 1: Exponents for Peak Spacing Fits}
\smallskip 
\settabs\+\hskip2.0truein&0.5\qquad&-1/3\qquad&-0.05\qquad&0.59\qquad&\cr 
\+&$h$&$\sg$&$x$&$y$&$z$\cr 
\smallskip 
\+&1&-1&0.06&0.59&0.91\cr
\+&1&-1/3&-0.05&0.59&1.21\cr
\+&0.5&-1&0.17&0.69&0.92\cr
\+&0.5&-1/3&0.05&0.69&1.18\cr 
\medskip 

We can also fit the change in peak spacing with $h$ as 
$$
\eqalign{\Dl l(\omo,\omt)_h&/\Dl l(\omo,\omt)_{h=1}=h^{-f(\omo)}
\cr f(\omo)&=0.30\omo^{-0.38}\,.\cr}\eqno(8) 
$$
This is good to better than 2\% for $h=0.5$.  Note that $f$ does not depend 
on $\omt$.  Also, 
the ratio of the peak spacings for the pure dust case ($\omo,\omo$) to the 
flat case ($\omo,1$) 
are independent of $h$, depending only on the equation of state, as can be 
seen from the quantity $y-x$ in Table 1. 

\medskip 
\centerline{\bf 5. Sensitivity to Cosmological Parameters} 
\medskip 
We now address how the information resident in the acoustic peaks traces 
and constrains the cosmological constituents.  See Hu \& White 
(1996), especially the excellent and comprehensive Section 5.6, for an 
outline of how the underlying physical process 
generating the matter perturbations influences the anisotropies. 
\medskip 
\leftline{\bf 5a. Determining Cosmological Density}
\smallskip 
As illustrated in the previous section, the dependence of the acoustic peak 
locations on the cosmological parameters is more complicated than initially 
believed.  In particular, the peak locations and spacing certainly do not 
completely determine the total density $\omt$; there is no unique value 
derivable without additional assumptions. 

To make this explicit, consider 
a putative determination of peak spacing at $\Dl l=350$.  Even if we 
restrict ourselves to only models with dust and cosmological constant, and 
with $h=1$ -- i.e.~we somehow have prior knowledge that Figure 1a is the 
appropriate one -- a horizontal line 
with $\Dl l=350$ intersects the curves at the 
following values: $(\omo,\omt)=$(0.48,0.48), (0.4,0.57), (0.3,0.69),  
(0.2,0.8), (0.1,0.92).  This means that a whole family of 
models is allowed, and these have total densities ranging from $\omt=0.48-
0.92$, when $\omo\ge0.1$.  This is very far from a definitive determination 
of the cosmological density and the fate of the universe, even 
with prior perfect knowledge of the types of cosmological constituents and 
of the Hubble constant.  (Even if one restricted to flat inflationary models, 
note that a measured $\Dl l$ of 290, say, could arise from $\omo=0.08$, 
$\omc=0.92$, $h=1$ or $\omo=0.52$, $\oms=0.48$, $h=0.5$ or $\omo=0.83$, 
$\omc=0.17$, $h=0.5$ -- quite a range of models.)  So the next ground based 
anisotropy detection 
hinting at a peak should not be expected to resolve or even significantly 
constrain our estimation of the density of the universe. 

Lest we be too pessimistic, we note that much additional information exists 
in the amplitudes of the multipole peaks and this will help constrain the 
cosmological parameters.  Analysis of these, however, is not so amenable 
to the quasianalytic arguments employed here and best relies on the excellent 
full numerical codes available. 

\medskip 
\leftline{\bf 5b. Constraining Subsidiary Cosmological Parameters}
\smallskip 
Turning from pessimism to optimism, let us consider what we can learn in 
that bright future (when MAP and Planck are aloft?) when superbly detailed 
CMB and other observations have fixed the major cosmological parameters 
like $\omo$, $\omt$, $h$, and $\sg$.  As mentioned in Section 3d a number of 
other cosmological quantities influence 
the acoustic peak spacing in a subsidiary role.  The two most promising 
for constraint are the number of neutrino species $N_\nu$ and the 
baryon-photon ratio $\eta$. 

Figure 2 illustrates the sensitivity of $\Dl l$ to both of these.  
As discussed in Section 3d, the number of neutrino species adds to the 
relativistic energy density, $\omr$, and hence affects the redshift of 
matter-radiation equality, $z_{eq}$ (which for three 
species is 21.6$\,\omo h^2\times\zd$).  The other key redshift is that 
of baryon-photon equality, $z_{eq(b,\gm)}$, a direct measure of $\eta$. 
The curves are labelled with the baryon loading parameter $\beta=
(3/4)z_{eq(b,\gm)}/z_{eq}=0.0158\,(\Om_bh^2/0.0125)\,(\omo h^2)^{-1}$. 

\nobreak\midinsert 
\centerline{
\epsfxsize=4.5truein
\epsfbox{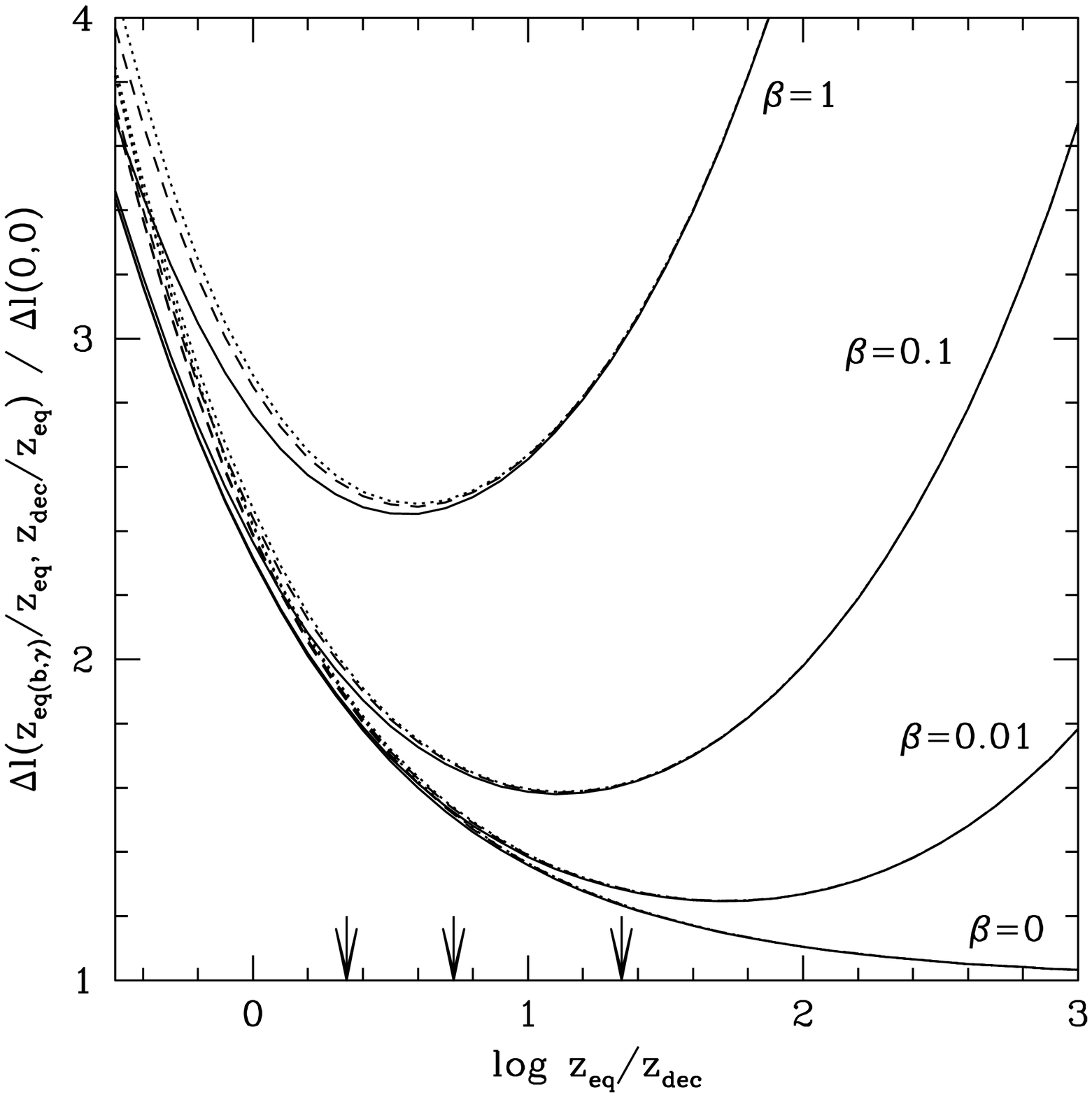}
}
\medskip\nobreak 
\noindent {\bf Figure 2:}\quad 
The peak multipole spacing is plotted relative to 
what it would be neglecting baryon loading and the radiation contribution 
to the expansion rate. This probes 
the influence of the baryon content of the universe, 
extra relativistic degrees of freedom, or anything that affects the ratio 
of the redshifts of matter-radiation equality and decoupling.  The arrows 
indicate the nominal values (with three light neutrino species) for 
universes with $\omo h^2=0.1,\,0.25,\,1$, from left to right.  The parameter 
$\beta$ is proportional to the baryon density.  Dotted 
curves neglect the effects on $r_a$ to show that the major influence is 
on the sound horizon, dashed curves include the effects on $r_a$ for 
$\omo=1$, and solid curves for $\omo=0.2$. 
\endinsert 
\bigskip 

As one increases the number of neutrino species, $N_\nu$, one decreases 
$z_{eq}$, and increases $\Dl l$ if the baryon loading is small.  What is 
plotted is the ratio of $\Dl l$ derived with some value of $z_{eq}$ 
compared to the asymptotic case where the universe at decoupling is 
completely matter dominated ($z_{eq}\gg \zd$).  So to find the effect 
of changing $N_\nu$ one simply takes the ratio of the multipole values 
for the two $N_\nu$'s considered.  For example, the value at the middle 
arrow (corresponding to three neutrino species in a $\omo h^2=0.25$ dust 
universe) is 1.54 for the $\beta=0.01$ curve.  Adding one light neutrino 
species 
decreases $\log(z_{eq}/\zd)$ by 0.055, bringing the value to 1.58, so 
$\Dl l$ increases by 2.5\%.  A reasonable fit is 
$$
\eqalign{(\Dl l)_{N_\nu}/(\Dl l)_3&=\left[1+0.135\,(N_\nu-3)\right]^
{g(\omo h^2)},\cr 
g(\omo h^2)&=0.07\,(\omo h^2)^{-0.54},\cr}\eqno(9)
$$
taking into account that $\beta$ changes with $\omo h^2$.  The 
effects tend to be in the 1-3.5\% range.  

One can also fix the number of neutrino species, and hence $\omr/\omo$ 
and $z_{eq}/\zd$, and consider changing the baryon-photon ratio $\eta$, 
which is proportional to $\beta$.  One has $\beta=[45\zeta(3)/2\pi^4] 
m_bT_\gm(\omr/\omo)\,\eta$, where $\zeta$ is the Riemann zeta function and  
$m_b=\rho_b/n_b$ the average baryonic particle mass.  So if all other 
parameters were determined, one 
could in principle measure the baryon-photon ratio, or equivalently the 
specific entropy of the universe, $s\sim \eta^{-1}$, from the shift of 
the acoustic peak spacing relative to its expected value for some fiducial 
$\eta$.  Note that even the helium abundance $Y$ enters, through 
the average baryon mass $m_b$.  Ignoring this last, tiny effect, one 
finds that $\Dl l\sim \eta^{0.11}$ for $\beta$ in the range 0.01-0.1, 
leading to a 8\% shift for a factor of two difference in $\eta$. 

Recall that for a dust, $h=1$ model $\Dl l\sim\omt^{-0.59}$ so a 3\% 
change in $\Dl l$ in this case, say, shifts $\omt$ by 5\%.  
The variation of subsidiary cosmological parameters leads only to small 
effects, but the Planck Surveyor should be able 
to pinpoint the multipole spacing to better than half a percent, so 
they are not wholly beyond detection. 

\bigskip 
\centerline{\bf 6. Conclusion} 
\medskip 
The physics behind the generation of the CMB acoustic peak anisotropies 
is well defined and simple.  However, common, 
asymptotic expressions for the distance behaviors and peak spacings are 
woefully inadequate.  Fortunately, there exist comprehensive numerical 
solutions, and this paper presents accurate analytic fits to the multipole 
peak locations taking into account the range of relevant 
cosmological parameters, elucidating the physical inputs.  While naive 
application of the peak location 
to determining the total density of the universe is doomed, full use of 
detailed anisotropy observations does indeed carry the hope of deriving 
not only the total density, but that of the individual components -- 
baryons, cosmological constant, etc.~-- and even such variables as the 
number of light neutrino species and the entropy of the universe. 

\medskip 
This work was supported by NASA grants NAG5-3525, NAG5-3922, and NAG5-4064. 
\vfill\break 
\centerline{\bf REFERENCES}
\medskip 
\parindent=0in 
Bond, J.R., Crittenden, R., Davis, R.L., Efstathiou, G., \& Steinhardt, 
P.J.~1994,\par 
\vskip-2pt 
\quad PRL, 72, 13

Bond, J.R., Efstathiou, G., and Tegmark, M.~1997, to appear in MNRAS, 
astro-ph/9702100 

Esmailzadeh, R., Starkman, G.D., \& Dimopoulos, S.~1991, ApJ, 378, 504 

Hu, W., Sugiyama, N., \& Silk, J.~1997, Nature 386, 37

Hu, W., \& White, M.~1996, ApJ, 471, 30 

Hu, W., \& White, M.~1997, in Proceedings of XXXIth Moriond Meeting: 
Microwave\par 
\vskip-2pt 
\quad Background Anisotropies, ed.~F.R.~Bouchet et al.~(Singapore: 
Editions Frontieres),\par 
\vskip-2pt
\quad 333 (astro-ph/9606140) 

Jones, B.J.T., \& Wyse, R.F.G.~1985, A\&A, 149, 144 

Kamionkowski, M., Spergel, D.N., \& Sugiyama, N.~1994, ApJ, 426, L57 

Kolb, E.W., \& Turner, M.S.~1990, The Early Universe, 
(Redwood City: Addison-Wesley) 

Linder, E.V.~1988a, A\&A, 206, 175 

Linder, E.V.~1988b, A\&A, 206, 190 

Linder, E.V.~1997, First Principles of Cosmology, (London: Addison-Wesley) 

Linder, E.V.~1998, ApJ, 497, in press, astro-ph/9707349 

Mattig, W.~1958, Astr.~Nachr., 284, 109 

Sakharov, A.D.~1966
\footnote{$^\spadesuit$}{Original publication 
in Russian in 1965.  This paper is also noteworthy 
for mentioning the first paper on the inflationary phase 
transition, published in the same volume: E.B.~Gliner 1966 (Russian 1965), 
Sov.~Phys.~JETP, 22, 378.}, 
Sov.~Phys.~JETP, 22, 241 

Turok, N.~1996, Phys.~Rev.~Lett., 77, 4138 

Zaldarriaga, M., Seljak, U., \& Bertschinger, E.~1997, submitted to ApJ, 
astro-ph/9704265 

Zaldarriaga, M., Spergel, D.N., \& Seljak, U.~1997, submitted to ApJ, 
astro-ph/9702157 
\bye